# Mitigating the Latency-Area Tradeoffs for DRAM Design with Coarse-Grained Monolithic 3D (M3D) Integration

Chao-Hsuan Huang, Ishan G Thakkar
Department of Electrical and Computer Engineering, University of Kentucky, Lexington, KY, U.S.A.
{chu276, igthakkar}@uky.edu

*Abstract*— Over the years, the DRAM latency has not scaled proportionally with its density due to the cost-centric mindset of the DRAM industry. Prior work has shown that this shortcoming can be overcome by reducing the critical length of DRAM access path. However, doing so decreases DRAM area-efficiency, exacerbating the latency-area tradeoffs for DRAM design. In this paper, we show that reorganizing DRAM cell-arrays using the emerging monolithic 3D (M3D) integration technology can mitigate these fundamental latency-area tradeoffs. Based on our evaluation results for PARSEC benchmarks, our designed M3D DRAM cell-array organizations can yield up to 9.56% less latency, up to 4.96% less power consumption, and up to 21.21% less energy-delay product (EDP), with up to 14% less DRAM die area, compared to the conventional 2D DDR4 DRAM.

## I. INTRODUCTION

DRAM is used as main memory in nearly all computing systems today. Since the emergence of DRAM, the continued scaling of the DRAM process technology has enabled a continuous decrease in the cost-per-bit of DRAM subsystems, as it has allowed a continuous increase in the DRAM cell density (i.e., DRAM cells per unit die area). The sustained scaling of the DRAM process technology has also allowed the fundamental DRAM access latency to decrease [1]. However, DRAM manufacturers have deliberately sacrificed the access latency benefits of DRAM process scaling to achieve lower cost-per-bit (i.e., greater cell density) for DRAM. As a result, the DRAM cell density has vastly improved by 128× in the past 20 years, but the DRAM latency has improved by 30% only [1]. The slower scaling of the DRAM latency has exacerbated the "Memory Wall" problem, due to the widening of the performance gap between the processor and DRAM subsystems even further. Breaking this "Memory Wall" is crucial for meeting the performance demands of the modern data-driven computing applications.

To alleviate the "Memory Wall" problem, the traditional approach for designing DRAM architectures has been to aim for a high cell density (i.e., high number of cells per DRAM die area) and low average access latency together. A common solution has been to enhance the microarchitecture of the conventional 2D DRAM modules to increase their memory access parallelism (e.g., [34], [35]). In contrast, in the past decade, several 3D-stacked DRAM architectures have emerged (e.g., [14], [28], [29]). In general, these 3D-stacked DRAM architectures employ *(i)* a larger number of banks per footprint area, and *(ii)* a shorter vertical memory access path (comprised of a through-silicon vias (TSVs) based interface), compared to the conventional 2D DRAM architectures. These benefits typically result in lower average memory access latency, coming at the extra cost of TSV-based 3D stacking of individual DRAM dies and other structural enhancements. The extra cost for 3D stacking is also because 3D stacking in fact decreases the cell density (i.e., cells per die area) for individual DRAM dies, in spite of it supporting greater number of total cells per footprint area. Despite this extra cost, several 3D-stacked DRAM architectures have already been adopted and standardized by the industry (e.g., Hybrid Memory Cube (HMC), High-Bandwidth Memory (HBM)), which corroborates the fact that the industry might be now ready to compromise the cell density per die and cost-per-bit of DRAM for lower access latency.

The key factor that drives the inherent tradeoff between the per-die cell density (i.e., cost) and latency in DRAM is the length of DRAM access path. In DRAM, a bit is stored as electrical charge on a capacitor-based cell. The small size of the capacitor can hold only a small amount of charge, which necessitates the use of a bulky sense amplifier (~150× larger than a DRAM cell [1]) to sense and amplify the charge to a full digital logic value for reading. To mitigate the large size of sense amplifiers (SAs), each SA is connected to many DRAM cells through a long wire called a bitline (which constitutes the critical DRAM access path). This design choice results in area-latency tradeoffs in DRAM. Longer bitlines (more DRAM cells connected to the bitline) require fewer SAs for given DRAM die capacity, reducing the die area (increasing the die's cell density) and cost-per-bit. But longer bitlines have increased electrical load, which increases the access latency. In contrast, shorter bitlines (fewer cells connected to the bitline) reduce the electrical load on the bitline, decreasing the access latency. But they require more SAs for a given die capacity, increasing the die area (reducing the die's cell density) and cost-per-bit. Despite of this inherent area-latency tradeoffs in DRAM, DRAM chips and architectures (e.g., [36], [37], [38]) with short bitlines (low latency) do exist; however, because of their lower per-die cell density (i.e., higher cost-per-bit) the industry has relegated them to specialized applications only such as high-end networking systems (e.g., [39]) that can tolerate a very high cost for a very low latency. *For more widespread adoption of the short-bitline DRAM architectures, the per-die cell density for such DRAM architectures needs to be increased, for which mitigating the fundamental latency-area tradeoffs for DRAM design is of paramount importance.*

To mitigate these latency-area tradeoffs for DRAM design, and consequently improve the per-die cell density for DRAM, we propose to use the emerging monolithic 3D (M3D) integration technology [2]. M3D technology enables sequential processing and integration of multiple tiers of logic circuits on the same die. To vertically connect various components located on different M3D tiers, the M3D-integraetd chips utilize monolithic inter-tier vias (MIVs) that are several orders of magnitude smaller in physical dimensions (~50nm×100nm) than TSVs (~1-3μm×10-30μm) [2]. Such small dimensions of MIVs enable M3D designs to facilitate nanoscale contact pitch for vertical interconnects, which in turn enables fine-grained (e.g., at the transistor-level and gate-level granularities) as well as coarse-grained (e.g., at the circuit-level and block-level granularities) partitioning of logical circuits across different M3D tiers. However, M3D fabrication process has some thermal-budget related shortcomings [24], due to which the performance of the transistors in the top tier and the interconnects in the bottom tier degrades [24], in a 2-tier M3D design. Despite these shortcomings, M3D integrated computing systems (e.g., [27]), network-on-chip

(NoC) architectures (e.g., [40]), and memory architectures (e.g., [21], [22], [25]), with up to 4 M3D tiers have been demonstrated to show significant performance benefits over the conventional designs. These outcomes highlight the excellent promise of the M3D integration technology.

In this paper, we utilize the M3D technology to reorganize the conventional DRAM cell-array at the coarse granularity. Our approach is to re-architect the traditional 1T1C (1-transistor 1-capacitor) DRAM design, which is totally different from the previous approach in [15] that demonstrated M3D-based extreme-density fabrication of the capacitor-less DRAM (i.e., double-gated floating body DRAM). Here we show for the first time that reorganizing the traditional 1T1C DRAM die (we consider DDR4 DRAM [13]) at the subarray-level granularity with the M3D technology can mitigate the inherent area-latency tradeoffs for DRAM design, in spite of suffering from performance degradation related to M3D integration. Our idea is to partition the sense-amplifiers and other peripherals on a different M3D tier from the tier with DRAM cell-arrays. We present two different M3D DDR4 DRAM designs, both with improved cell density (die area) and access latency, compared to the baseline 2D DDR4 DRAM of the same capacity.

Our key contributions in this paper are summarized below.

- We evaluate latency-area tradeoffs for various DDR4 DRAM organizations with different local bitline lengths and find that reducing the length of local bitlines can lead to increased DRAM access latency in some cases.
- To achieve reduced access latency with shorter bitlines, we re-organize the cell-array of the commodity 2D DDR4 DRAM [13] using the coarse-grained M3D integration technology; We come up with two different designs of M3D DDR4 DRAM with reduced access latency;
- We present the subarray-level bank layouts as well as the latency, area, and energy analysis (based on SPICE and other circuit-level simulations) for our designed M3D DDR4 DRAMs;
- We show that our designed M3D DDR4 DRAM architectures achieve relaxed four bank activation window (tFAW) timing, which improves their bank-level parallelism to further reduce their average access latency;
- We evaluate our designed M3D DDR4 DRAM architectures using Gem5 [10] based full-system simulations with PARSEC benchmarks [11], and compare their performance, power, and energy-efficiency with the conventional 2D DDR4 DRAM.

## II. BACKGROUND ON DRAM STRUCTURE AND OPERATION

### A. Background on DRAM Chip Structure

A DRAM chip typically employs a hierarchical cell-array organization, which is briefly illustrated in Fig. 1. A DRAM chip (Fig. 1(a)) is divided into multiple banks (Fig. 1(b)), each of which is further divided into multiple subarrays (Fig. 1(c)). Every subarray in the bank is connected to the bank I/O via global bitlines and global sense amplifiers (Fig. 1(c)) [1]. A subarray contains multiple tiles (32 tiles in our considered example), all of which work in tandem (Fig. 1(d)). Each tile typically contains 512 cells in a vertical bitline and 512 cells in a horizontal wordline, forming a 512×512 cell-array structure with 512 local bitlines and 512 local wordlines. A cell is the smallest unit in the hierarchy, which consists of one transistor that connects to a local bitline, a local wordline and a capacitor (Fig. 1(e)). The amount of charge stored in the capacitor represents the stored bit value as '0' (if, not charged) or '1' (if, fully charged). To read the stored bit value, the sense amplifier (SA) located at the end of each bitline (Fig. 1(e), 1(d)) senses the charge stored in the cell capacitor and distinguishes the stored '0' from '1'.

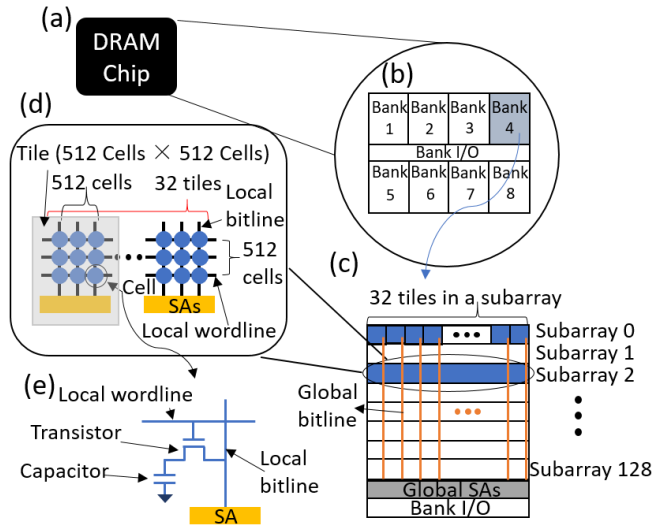

Fig. 1: Schematic structures of (a), (b) a DRAM chip, (c) a DRAM bank, (d) a DRAM subarray, and (e) DRAM cell. SAs: Sense Amplifiers.

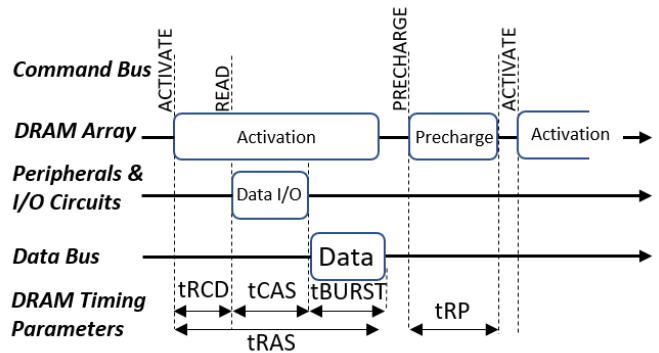

Fig. 2: Three phases of DRAM operation and related timing parameters.

### B. DRAM Operation Commands and Timing Constraints

In general, as shown in Fig. 2, the DRAM operation (i.e., DRAM cell-array access) can be broken down into three distinct phases: (i) activation, (ii) data I/O, and (iii) precharging. In addition, the DRAM operation also involves periodic refresh phases (not shown in Fig. 2), which are generally scheduled as a series of multiple dummy activation and precharging phases [34]. From Fig. 2, activation and precharging phases occur entirely within the subarray, whereas the data I/O phase occurs in the peripherals and I/O circuitry. Moreover, the data I/O phase overlaps with the activation phase. In addition to the above-mentioned phases, a DRAM subarray can also be in the dormant phase during which no dynamic activity occurs in the subarray. Each activity phase of DRAM has its own timing constrains to ensure error-free DRAM operation. To meet these timing constraints, a DRAM subsystem employs a DRAM controller that typically resides on the host processor die. To navigate through the three activity phases, the DRAM controller uses three basic commands: ACTIVATE (ACT), READ/WRITE, and PRECHARGE (PRE). In this section, we briefly explain the timing constraints (Fig. 2) and internal DRAM operation (Fig. 3) related to all three DRAM phases and associated commands.

**Dormant Phase:** Fig. 3(a) shows a representative structure of a DRAM cell and its associated local bitline in the *dormant* state. Here the access transistor is shown as a switch and a capacitor $C_{LBL}$ is shown connected to the local bitline to represent the bitline's parasitic capacitance. In addition, the cell capacitor ($C_C$) is shown to be fully charged to $V_{DD}$ voltage level to represent that bit '1' is stored in it. In the *dormant* state, $C_{LBL}$ is generally pre-charged at $0.5V_{DD}$. Moreover, a global bitline is also shown with its parasitic capacitor $C_{GBL}$. The global bitline is shown to connect to a SA I/O at one end and to a global SA at the other end. $C_{GBL}$ is also precharged to $0.5V_{DD}$, although the operation of a global bitline is a little different from the local bitline operation, as will be clear later in this section.

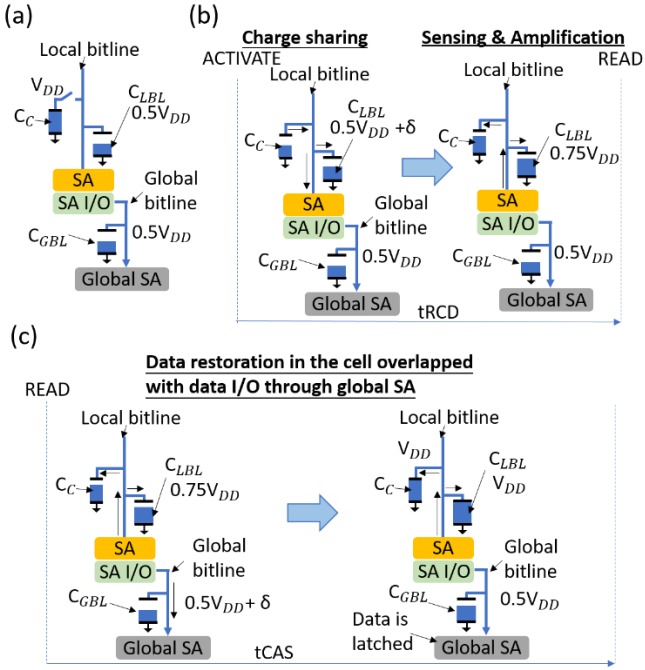

**Fig. 3:** Illustrations of the DRAM cell operation during (a) dormant phase, (b) activation phase, and (c) data I/O phase.

**Activation Phase:** The activation phase starts when the DRAM controller issues the ACT command. All the local bitlines in the target subarray have to be precharged to $0.5V_{DD}$ before the ACT Command can be issued. Upon issuance of the ACT command, a target row is selected in the target subarray through the issued row select address. All the access transistors in the selected row are turned on to connect their respective cell capacitors to the local bitlines. Establishment of this connection makes every cell capacitor $C_C$ in the selected row to share its charge (or the lack thereof) with the $C_{LBL}$ of its corresponding local bitline. This sharing of charge between $C_C$ and $C_{LBL}$ (Fig. 3(b)) results in the $C_{LBL}$ voltage to change from the dormant value of $0.5V_{DD}$ to either $0.5V_{DD}+\delta$ if the initial voltage on $C_C$ was $V_{DD}$ (i.e., '1' was stored in $C_C$) (Fig. 3(b)) or $0.5V_{dd}-\delta$ if the initial voltage on $C_C$ was 0V (i.e., '0' was stored in $C_C$). During this charge sharing stage of the activation phase, the charge on $C_C$ is modified (i.e., the voltage on $C_C$ is modified and data is lost) as it is shared with $C_{LBL}$ (Fig. 3(b)). But this data loss is only temporary as the charge/voltage on $C_C$ is restored during the sensing and amplification stage of the activation phase, as discussed next.

After the charge sharing stage is complete, the SA connected with each local bitline ($C_{LBL}$) is enabled. The SA senses the change in $C_{LBL}$ voltage from its dormant value of $0.5V_{DD}$ and "amplifies" that change towards the full swing (i.e., $0.5V_{DD}+\delta$ is "amplified" towards $V_{DD}$, whereas $0.5V_{DD}-\delta$ is "amplified" towards 0). For that, the SA either injects or withdraws charge from $C_{LBL}$. After a latency of tRCD (Fig. 3(b) and Fig. 2), midway through amplification, enough charge has been injected (or withdrawn) such that the $C_{LBL}$ voltage reaches a threshold state of $0.75V_{DD}$ (or $0.25V_{DD}$). At this point, data is considered to have been "copied" from the cell to the SA. In other words, the $C_{LBL}$ voltage is now close enough to $V_{DD}$ (or 0) for the SA to detect a '1' (or '0') and transfer it to the SA I/O, allowing READ and WRITE commands to be issued. Until this stage, the SA I/O, global bitline, and global SA remain dormant (Fig. 3(b)). After this stage, they are activated (Fig. 3(c)) and the READ/WRITE phase starts in the peripherals and I/O circuitry (Fig. 2), while the activation phase continues inside the subarray (Fig. 2). Eventually during the continued activation phase, the $C_{LBL}$ voltage and $C_C$ voltage are fully amplified to $V_{DD}$ or 0. Only at this point is the charge and data in the cell $C_C$ fully restored to its original value. The latency to reach this restored state is tRAS (Fig. 2). Thus, the values of tRCD and tRAS depend on how fast $C_{LBL}$ and $C_C$ can be charged (or discharged).

**Data I/O Phase:** Upon the issuance of the READ/WRITE command, the data I/O phase starts and the SA I/O starts pushing out the data to global SA (Fig. 3(c)). The SA I/O does that by injecting (or withdrawing) charge from the global bitline capacitor ($C_{GBL}$) [34]. As a result, the voltage across $C_{GBL}$ is perturbed to $0.5V_{DD}+\delta$ (or $0.5V_{DD}-\delta$). The global SA senses and amplifies this perturbation to latch the data value (Fig. 3(c)) before relaying it to the bank I/O (not shown in the figure). However, unlike the local SA, the global SA does not need to "amplify" the $C_{GBL}$ voltage towards the full swing. This is because while injecting (or withdrawing) charge into $C_{GBL}$, the local SA I/O does not lose the original data value. Thus, this is how the operation of local bitlines differ from the operation of global bitlines. The latency from the local SA I/O starts pushing out the data until the global SA latches the data is tCAS (Fig. 2). Thus, tCAS latency depends on $C_{GBL}$ value.

**Precharging Phase:** Precharging phase starts right after the activation phase (Fig. 2). After the activation phase is complete, the $C_{LBL}$ voltage for all local bitlines in the subarray will either be $V_{DD}$ or 0. At this stage, the PRE command is issued that decouples the DRAM cells from their respective local bitlines and makes the precharge units (not shown in Fig. 3) that are typically the part of SAs to withdraw (or inject) current from (or into) the bitline capacitor $C_{LBL}$ until the $C_{LBL}$ voltage ends up being $0.5V_{DD}$. The latency for this phase is tRP (Fig. 2).

*In summary*, the critical DRAM timing/latency parameters tRCD, tCAS, and tRP depend on the local bitline parasitic capacitance $C_{LBL}$ and the global bitline parasitic capacitance $C_{GBL}$, and the values of $C_{LBL}$ and $C_{GBL}$ depend on the lengths of the local and global bitlines respectively. Typically, longer bitlines have higher $C_{LBL}$ and $C_{GBL}$ values, which results in longer delay for the charging and discharging of the bitlines, causing longer tRCD, tCAS, and tRP latencies for the DRAM. This observation touts having shorter bitlines as the fundamental approach for reducing the DRAM latency parameters tRCD, tCAS, and tRP.

## III. LATENCY-AREA TRADEOFFS FOR 2D DRAMs

As discussed in Section II-B, having shorter bitlines is the fundamental approach for reducing the tRCD, tCAS, and tRP latencies of DRAM. However, from [1], reducing the length of local bitlines to reduce tRCD and tRP comes at the cost of increased die area and decreased per-die cell density. This is because having shorter local bitlines entails a smaller number of cells that are shared per SA, which in turn requires more SAs for the given die capacity, increasing the die area and reducing the per-die cell density. However, in [1], such latency-area tradeoffs are analyzed for only tRCD and tRP parameters. In contrast, in this paper, *we analyze the latency-area tradeoffs for tCAS parameter for the first time, and non-intuitively find that shortening the local bitlines can lead to requiring longer global bitlines for the same bank and die capacities, to yield higher tCAS latency. We also find that the higher tCAS latency can offset the benefits of shorter local bitlines to render higher DRAM access latency.*

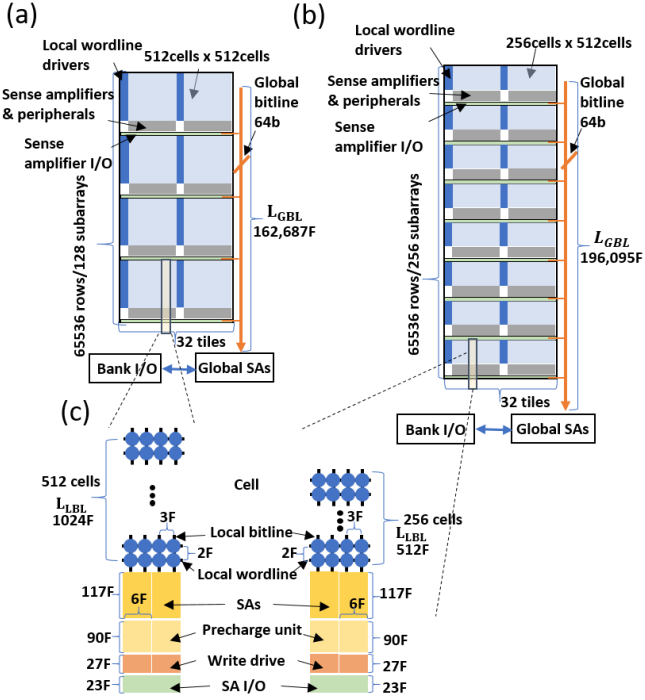

Fig. 4: Illustration of two example bank organizations of the folded-bitline 2D DDR4 DRAM with (a) 512 cells per local bitline (DDR4-512) and (b) 256 cells per local bitline; (c) shows the enlarged schematics of the DDR4-512 and DDR4-256 subarrays. SAs: sense amplifiers.

To understand our observation, consider Fig. 4 that shows the schematic organizations of a 2D DRAM bank (DDR4 [13]) for two different local bitline lengths, i.e., 512 cells and 256 cells per local bitline. Henceforth, the bank organizations with 512 cells per local bitline (Fig. 4(a)) and 256 cells per local bitline (Fig. 4(b)) are respectively referred to as DDR4-512 and DDR4-256 organizations. The DRAM chip we have considered for this analysis is of 1Gb in size with total 8 banks of 128Mb size each. Each bank has total 65536 rows with each row horizontally split across 32 tiles (i.e., total 32 tiles per subarray). Moreover, there are total 512 DRAM cells in every local wordline. In the bank organizations shown in Fig. 4, we have indicated the lengths of global and local bitlines (Fig. 4(a) and 4(b)), as well as the dimensions of the area covered by other critical peripherals such as SAs, SA I/O, precharge units, and write drivers (Fig. 4(c)). We evaluated the bitline lengths and peripherals' area dimensions using the CACTI [33] based DRAM model for 22nm, considering the folded bitline architecture from [32]. We select 22nm technology node for our evaluations because the DRAM technology roadmap beyond 22nm node in terms of the available DRAM cell structures and fabrication options is not standardized across different DRAM manufacturers.

For the DDR4-512 organization in Fig. 4(a), the length of the local bitlines ($L_{LBL}$) is 1024F (indicated in Fig. 4(c)), where F is the minimum feature size for a given technology node. Having 512 cells per local bitline renders total 128 subarrays in the bank (Fig. 4(a)). Therefore, the global bitlines in the DDR4-512 organization must span total 128 subarrays (=65536 total rows / 512 rows per subarray) and their respective bitline peripherals areas, which makes the length of the global bitlines ($L_{GBL}$) to be 162, 687F (Fig. 4(a)). On the other hand, for the DDR4-256 organization shown in Fig. 4(b), $L_{LBL}$ is 512F which is half of the $L_{LBL}$ value for the DDR4-512 organization. This is obviously because the DDR4-256 organization has 2× less number of cells per local bitline. Moreover, we evaluated the tRCD latency for the DDR4-512 and DDR4-256 organizations to be 6.8ns and 5.0ns, respectively (based on SPICE simulations discussed in Section IV and V). Thus, the bank organization with shorter local bitline length (i.e., DDR4-256 in Fig. 4(b)) has lower tRCD latency. However, from Fig. 4(b), owing to the 2× shorter local bitlines, the DDR4-256 organization has 2× more subarrays, which results in longer global bitlines with $L_{GBL}$ to be 196,095F. Due to the longer $L_{GBL}$ the DDR4-256 organization achieves longer tCAS latency of 12.0ns, compared to the 10.3ns tCAS latency for the DDR4-512 organization.

However, having a lower tCAS latency or tRCD latency alone may not always result in lower DRAM access latency, as the DRAM access latency depends on different combinations of DRAM latency parameters, depending on the utilized page (or row buffer) management policy. From [28], the close-page policy works better for modern learning-based applications, due to the low spatial locality offered by the random access patterns of such applications. Therefore, instead of comparing the individual latency parameters for different bank organizations, more insightful comparison will be of the close-page access latency. Typically, the close-page access latency for DRAM is given by tRCD+tCAS+tBURST, where tBURST is 4 cycles (i.e., 4ns at 1GHz) for our considered DDR4 based DRAM organizations. Accordingly, the close-page access latency for our DDR4-512 and DDR4-256 organizations become 21.1ns and 21.0ns, respectively. Thus, having 2× shorter local bitlines for DDR4-256 compared to DDR4-512 does not translate into any significant access latency benefit for DDR4-256 (benefit of only 0.1ns, which is negligible for the typical DDR4 clock rate of 1GHz-3GHz [13]), in spite of DDR4-256 rendering lower tRCD value. Moreover, the tRCD benefits for DDR4-256 comes at the cost of extra area consumption, as DDR4-256 it has 2× more subarrays than DDR4-512. Therefore, it is not clear if further reducing $L_{LBL}$ below the $L_{LBL}$ value for DDR4-256 can provide additional benefits in terms of die area and/or close-page access latency.

This outcome motivated us to evaluate the latency-area tradeoffs for different bank organizations having different local bitline lengths. We evaluated the die area, tRCD and close-page access latencies for bank organizations with 128 cells per local bitline (DDR4-128), 64 cells per local bitline (DDR4-64), and 32 cells per local bitline (DDR4-32), using our CACTI and SPICE based DDR4 model discussed earlier. The results of our evaluation are plotted in Fig. 5. Fig. 5 also plots results for our proposed M3D organizations, which will be discussed in the next section. From Fig. 5, compared to the DDR4-256 organization, the tRCD latency reduces and die area increases for the DDR4-128 to DDR4-64 organizations. However, the benefits in tRCD latency saturate at the DDR4-64 configuration, that is, tRCD does not reduce beyond DDR4-64 for DDR4-32, even with 2× reduction in number of cells per local bitline and ~1.7× increase in the die area. In contrast, the benefits in close-page access latency saturate much earlier for the DDR4-256 configuration. For the DDR4-128, DDR4-64, and DDR4-32 configurations, both close-page access latency and die area increase.

Thus, it is clear from these findings that reducing the local bitline length below 256 cells is not beneficial in terms of die area (per-die cell density) and access latency, unless the length of the global bitlines can also be reduced in concurrence without incurring any extra die area cost. To this end, it seems intuitively encouraging to reduce the bank size and increase the bank count per DRAM die to concurrently reduce the global bitline length. However, doing so cannot come without significantly harming the per-die cell density of DRAM. Therefore, to reduce the lengths of both the local and global bitlines of DRAM banks without reducing the per-die DRAM cell density, we propose to reorganize DRAM banks using the monolithic 3D integration (M3D) technology, as discussed next.

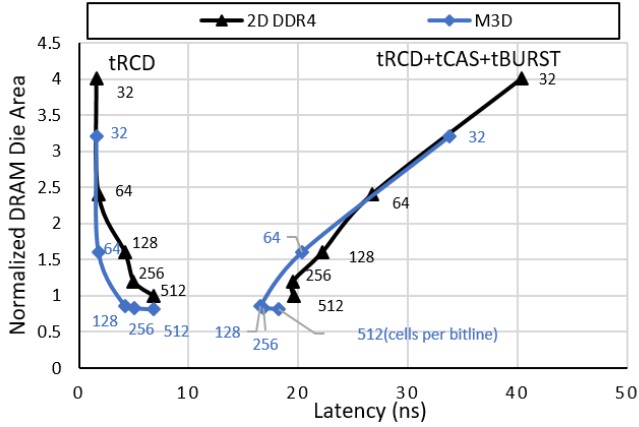

**Fig. 5: Normalized DRAM die area versus tRCD and close-page access latency (tRCD + tCAS + tBURST) for various local bitline lengths (cells per local bitline) for the 2D DDR4 DRAM. tRCD is row-to-column access delay, tCAS is column access strobe latency, and tBURST is data burst duration.**

## IV. REORGANIZING DRAMS WITH M3D INTEGRATION

### A. Monolithic 3D Integration Technology

Before going into the specifics of how we use the monolithic 3D (M3D) integration technology to reorganize DRAMs, we first provide a brief overview of the M3D technology. M3D technology enables sequential processing and integration of multiple tiers (mostly up to two tiers) of logic circuits on the same die. To vertically connect various components located on different M3D tiers, the M3D-integraetd chips utilize monolithic inter-tier vias (MIVs) that are several orders of magnitude smaller in physical dimensions (~50nm×100nm) than TSVs (~1-3µm×10-30µm) [40]. Such small dimensions of MIVs enable M3D designs to facilitate nanoscale contact pitch for vertical interconnects, which cannot be achieved using the traditional through-silicon via (TSV) based 3D-stacking technology. Moreover, an MIV has 10Ω resistance and 0.2fF capacitance, *which enables vertical routing of connections using MIVs with negligible overheads of parasitic loading*. More details on the M3D integration technology can be found in [3].

The disadvantage of M3D integration is that the second M3D tier is sequentially grown on top of the back-end-of-line metal interconnects of the active bottom tier. To ensure that the bottom tier metal interconnects withstand the high temperatures during the fabrication process of the second tier, tungsten is used as the interconnect metal on the bottom tier instead of copper. However, tungsten has 2× resistivity than copper, which harms the performance of tungsten interconnects on the bottom tier. On the other hand, the manufacturing process for the sequential integration of the second tier is not yet perfect, which degrades the transistor performance in the second/top tier by 10~20%. To mitigate this tier degradation issue, we employ an established workaround from [24], to make the best use of the M3D technology for designing better performing DRAM organizations, as discussed next.

### B. Design of Monolithic 3D (M3D) DRAMs

For reorganizing DRAMs with M3D technology, our idea is to partition the sense-amplifiers (SAs) and other peripherals on a different M3D tier from the tier with DRAM cell-arrays. But the tier degradation issue of M3D integration raises an important question: Should the SAs and other peripherals be placed on the bottom tier or top tier? As the SAs and other peripherals are mainly logic-intensive circuits and DRAM cell-arrays are mainly interconnects-intensive circuits (because of the utilized bitlines and wordlines), the SAs and peripherals (DRAM cell-arrays) are more prone to tier degradation effects if they are placed on the top (bottom) tier. Therefore, to avoid such performance degradation, we decided to place the SAs and other peripherals (e.g., write drivers, precharge units, SA I/O, local wordline drivers, address decoders) on the bottom tier, and the DRAM cell-arrays (including the DRAM interconnects such as bitlines and wordlines) on the top tier, in our designed M3D bank organizations. We route the connections of the SAs and other peripherals on the bottom tier to the DRAM interconnects on the top tier using MIVs and tier-specific metal-via stacks. Next we explain our designed M3D bank organizations in detail.

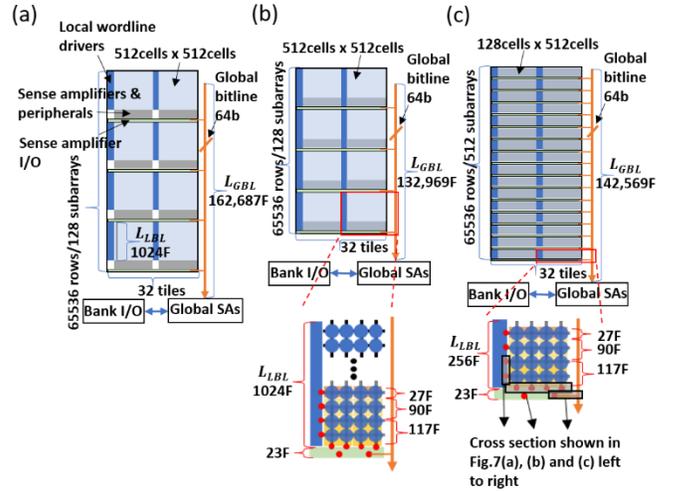

**Fig. 6: Illustration of three example bank organizations of the folded-bitline DRAM; (a) 512 cells per local bitline 2D DDR4 DRAM (DDR4-512), (b) 512 cells per local bitline M3D DRAM (M3D-512), and (c) 128 cells per local bitline M3D DRAM (M3D-128). Although the local/global address decoders are not shown, they are placed on the bottom tier.**

**M3D DRAM Bank Organizations:** Fig. 6 shows two schematic organizations of an M3D DRAM bank for two different local bitline lengths, i.e., 512 cells and 128 cells per local bitline, and compares them with the DDR4-512 organization from Fig. 4 (reproduced as Fig. 6(a)). Henceforth, the M3D bank organizations with 512 cells per local bitline (Fig. 6(b)) and 128 cells per local bitline (Fig. 6(c)) are respectively referred to as M3D-512 and M3D-128 organizations. From Fig. 6, for the M3D organizations, we place the SAs, precharge units and write drivers on the bottom tier of the die right underneath the DRAM tiles that are placed on the top tier. Local SA I/O is also located on the bottom tier, but not directly underneath the DRAM tiles; we make this design choice to ensure that the vertical routing of interconnects from the bottom tier and top tier is feasible. Placing the SAs and peripherals directly underneath the DRAM tiles saves die area to increase per-die cell density. Moreover, it also helps in reducing the global bitline length without incurring the extra area overhead of dividing the DRAM cell-array into multiple banks. For example, for the M3D-512 organization in Fig. 6(b), placing SAs and peripherals underneath the DRAM tiles saves 234F (117F for SAs + 90F for precharge units +

27F for write drivers; enlarged inset of Fig. 6(b) with dimensions from Fig. 4) of the global bitline length $L_{GBL}$ per subarray, yielding $L_{GBL}$ to be 132,969F for M3D-512, compared to $L_{GBL}$ of 162,687F for DDR4-512 (Fig. 6(a) and Fig. 4(a) & 4(c)) that has the same local bitline length $L_{LBL}$ of 1024F as M3D-512. As a result of reduced $L_{GBL}$, M3D-512 achieves reduced tCAS of 8.9ns, compared to tCAS of 10.3ns for DDR4-512. Moreover, we evaluate that the area of a 128Mb M3D-512 bank is 3.2mm$^2$, which is significantly less than the 3.9mm$^2$ area of a 128Mb DDR4-512 bank. This outcome motivated us to evaluate the latency-area tradeoffs for different M3D bank organizations having different local bitline lengths, as discussed next.

**Latency-Area Tradeoffs for M3D DRAM Organizations:** In addition to the M3D-512 organization, we also evaluated the die area, tRCD and close-page access latencies for four other M3D bank organizations with 256 cells per local bitline (M3D-256), 128 cells per local bitline (M3D-128), 64 cells per local bitline (M3D-64), and 32 cells per local bitline (M3D-32), using our CACTI and SPICE based DRAM model discussed in Section V. The results of our evaluation are plotted in Fig. 5. From the figure, the tRCD and close-page access latency curves for the M3D organizations are closer to the origin than the curves for the DDR4 organizations, which indicates that the M3D organizations relax the fundamental latency-area tradeoffs for DRAM design. Along the same lines, the pinnacle of the benefits of M3D integration is achieved for the M3D-128 organization (Fig. 6(c)), for which $L_{GBL}$ of 142,569F and $L_{LBL}$ of 256F are achieved (Fig. 6(c)). These values of $L_{GBL}$ and $L_{LBL}$ are ~1.14× and 4× less respectively than the $L_{GBL}$ and $L_{LBL}$ values for DDR4-512. Moreover, we evaluate that the area of a 128Mb M3D-128 bank is 3.4mm$^2$, which is significantly less than the 3.9mm$^2$ area of a 128Mb DDR4-512 bank. *These results corroborate the excellent capabilities of the M3D technology in mitigating the fundamental latency-area tradeoffs for DRAMs, to achieve simultaneous benefits in DRAM access latency and per-die cell density.*

**Implementation Overheads for M3D DRAM Organizations:** As mentioned earlier, the feasible implementation of our proposed M3D DRAM organizations requires efficient and low-overhead routing of vertical connections between the peripherals on the bottom tier and cell-arrays on the top tier. In our proposed M3D DRAM organizations, this vertical routing of interconnects is achieved using MIVs and tier-specific metal-via stacks. For instance, Fig. 7(a) shows the cross-section of the vertical interconnects between the local wordline drivers on the bottom M3D tier and the local wordlines on the top M3D tier. Evidently, each vertical connection includes one M1-M5 metal-via stack and an MIV. Similarly, Fig. 7(b) shows how the bottom tier SAs are connected to the top tier bitlines (typically implemented in M1 of the top tier), and Fig.7(c) demonstrates how the SA I/O on the bottom tier is connected to a local data line (M1) and global bitline (M3) on the top tier. We extract the parasitic resistance and capacitance values for the vertical interconnects from [31] and find these values to be 0.23fF and 20Ω for the worst-case scenario (i.e., highest parasitic loading) shown in Fig. 7(c). In addition to this parasitic resistance and capacitance of the vertical interconnects, our M3D organizations also suffer from the performance degradation of the DRAM cell access transistors placed on the top tier. We evaluate this degradation in terms of $I_{ON}$-$I_{OFF}$ characteristics using the methods from [24]. We incorporate the vertical interconnects' parasitic values and the degraded access transistors' $I_{ON}$-$I_{OFF}$ characteristics in our LTSpice model from [30], to evaluate their impact on various DRAM latency parameters such as tRCD and tRP. Fig. 8 shows the results of our LTSpice simulations for tRCD parameter extraction for the DDR4-512, M3D-512, and M3D-128 organizations. As discussed earlier, both DDR4-512 and M3D-512 have the same value of 1024F for $L_{LBL}$. From Fig. 8, even with the addition of parasitic overheads of vertical interconnects and performance degradation of the access transistor, tRCD latency for M3D-512 hardly changes significantly compared to the tRCD latency for DDR4-512. *From these findings, we can conclude that M3D integration incurs negligible overhead for our proposed M3D DRAM organizations.*

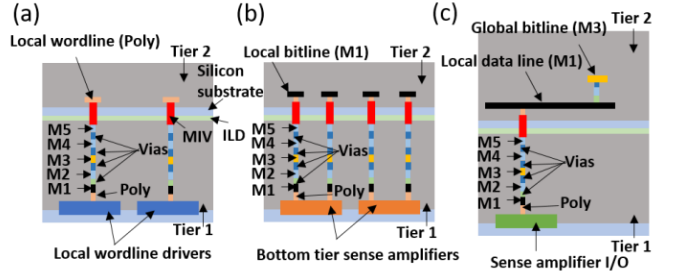

**Fig. 7:** Illustration of the vertical interconnects' cross-sections between (a) local wordline drivers and local wordlines, (b) sense amplifiers (SAs) and local bitlines, and (c) sense amplifier (SA) I/O and local data line & global bitline, for our proposed M3D DRAMs. ILD: Inter Layer Die electric; MIV: Monolithic Inter-tier Vias. Although the local/global address decoders are not shown, they are placed on the bottom tier.

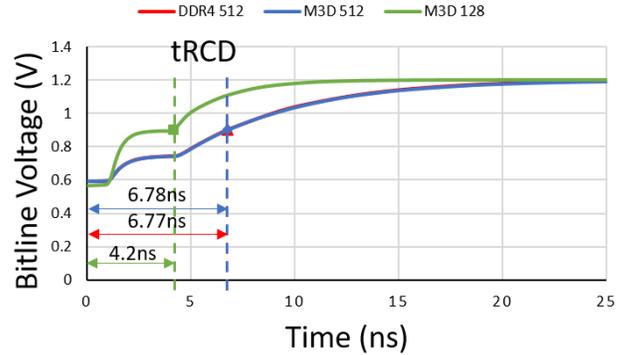

**Fig. 8:** Results of LTspice simulations for tRCD extraction for the DDR4-512, M3D-512 and M3D-128 organizations.

## V. AREA, TIMING, AND ENERGY ANALYSIS

### A. Area, Timing, and Energy Analysis

We modeled various DRAM organizations for 22nm technology node using CACTI [33]. Each DRAM cell consumes $6F^2$ area, while the height and pitch of a SA are 117F and 6F respectively. We evaluate the lengths of local and global bitlines also using CACTI based models of various DDR4 and M3D organizations. For M3D organizations, we hide the area consumed by the SAs and other peripherals, to come up with bank and DRAM die area. We extract energy values from CACTI based models as well. Moreover, to evaluate various DRAM latency parameters and close-page access latency, for the area and latency analysis, we use the sense amplifier with DRAM subarray bitline model from [30] in LTspice [7]. The model from [30] is for 45nm, so we scale it to 22nm following the standard guidelines scaling wires and interconnects in CMOS technologies. Our extracted modeling parameters are listed in Table 1 for various DDR4 and M3D DRAMs.

### B. Effect of M3D Integration on tFAW Timing

Because an ACT command consumes a lot of power/current [34], DRAM standards define a timing parameter to constrain the activity rate of DRAM so that ACT commands do not over-stress the power delivery network (PDN) [34]. The parameter is called the four activate window (tFAW) that defines the length of a rolling window during which a maximum of four ACT commands can be in progress. From the Micron DRAM power model [8] and the 8Gb DDR4 datasheet [13], we extract the tFAW time our DDR4-512 organization to be 35.8ns (Table 1). From Table 1, as the activation energy for the M3D-512 and M3D-128 organizations are less than the activation energy for the DDR4-512 organization, there is an opportunity to reduce the tFAW time for the M3D-512 and M3D-128 organizations. This is because, due to their lower activation energy consumption, the M3D-512 and M3D-128 organizations can be expected to put less current load on their PDN during ACT commands. Therefore, we utilize the method presented in [28] to scale down the tFAW timings for the M3D-512 and M3D-128 organizations. Table 1 lists our evaluated tFAW timings. Evidently, from tFAW of 35.8ns for the DDR4-512 organization, the tFAW values for the M3D-512 and M3D-128 organizations scale down to 35.3ns and 14.4ns respectively. As a result, the M3D-512 and M3D-128 organizations have increased inherent access parallelism compared to the DDR4-512 organization.

TABLE 1. MODELING PARAMETERS FOR VARIOUS DDR4 AND M3D DRAM ORGANIZATIONS.

|  | 2D DDR4 | M3D 512 | M3D 128 |
|---|---|---|---|
| Ranks | 1 | 1 | 1 |
| Banks | 8 | 8 | 8 |
| Page size | 16kb | 16kb | 16kb |
| Cells per bitline | 512 | 512 | 128 |
| Timing parameters (ns) | | | |
| tRCD | 6.77 | 6.78 | 4.2 |
| tCAS | 10.29 | 8.96 | 9.82 |
| tRP | 9.58 | 9.6 | 4.04 |
| tRC | 26.64 | 25.34 | 18.05 |
| tFAW | 35.8 | 35.3 | 14.4 |
| tREFI | 7800 | 7800 | 7800 |
| Per access energy values (nJ) | | | |
| Activation Energy | 0.59 | 0.58 | 0.24 |
| Read Energy | 1.1 | 0.94 | 1.05 |
| Write Energy | 1.1 | 0.94 | 1.05 |
| Refresh Energy | 35.22 | 32.51 | 23.23 |
| Area analysis | | | |
| Subarray (mm$^2$) | 0.031 | 0.025 | 0.007 |
| Bank (mm$^2$) | 3.926 | 3.209 | 3.42 |
| #MIVs per bank | 0 | 5,243,008 | 14,680,576 |
| MIV area per bank (mm$^2$) | 0 | 0.01 | 0.029 |
| Subarray height | 1281F | 1047F | 279F |
| Local bitline length | 1024F | 1024F | 256F |
| Local bitline resistance | 20000Ω | 20010Ω | 5010Ω |
| Local bitline capacitance | 72fF | 72.2fF | 18.2fF |

## VI. SIMULATION SETUP AND RESULTS

We performed trace-driven simulations using NVmain [9] to compare the power and energy-delay product values for our considered DRAM organizations. We consider DDR4-512, M3D-512, and M3D-128 organizations for system-level comparison. We also perform full-system simulations in Gem5 [11], to evaluate cycles per instruction (CPI) and average latency results. We used the PARSEC benchmarks [10] for the analysis, the trace files were extracted from detailed cycle-accurate simulations using GEM5 [11]. The configuration of GEM5 for both trace-driven and full-system simulations is shown in Table 2. We considered 11 different applications form the PARSEC suite: Blackscholes, Bodytrack, Canneal, Dedup, Facesim, Ferret, Freqmine, Streamcluster, Swaptions, Vips, and X264. For the trace-driven simulations, we ran each PARSEC benchmark for a "warm up" period of one billion instructions and captured memory access traces form the subsequent one billion instructions extracted. For the full-system simulations, we run PARSEC benchmarks in their critical regions of interest (ROIs) in Gem5. We use parameters from Table 1 to model the DDR4-512, M3D-512, and M3D-128 organizations in Gem5 and NVMain.

TABLE 2. GEM5 CONFIGURATION FOR TRACE-DRIVEN AND FULL-SYSTEM SIMULATIONS.

| Number of Cores | 4 | L2 Coherence | MOESI |
|---|---|---|---|
| L1 I Cache | 32KB | Frequency | 2 GHz |
| L1 D Cache | 32KB | Issue policy of cores | OoO (4 issue) |
| Shared L2 Cache | 2MB | # Memeory Controllers | 1 |
| ISA/OS | ALPHA | Cache Associativity | 4-way (L1); 8-way (L2) |

Fig. 9(a) shows system-level cycle per instruction (CPI) values for our considered DRAM organizations across PARSEC benchmarks. Compared to the baseline DDR4-512, M3D-512 and M3D-128 organizations yield about 0.54% and 3.74% lower system CPI respectively. Similarly, Fig. 9(b) shows average access latency values. Compared to the baseline DDR4-512, M3D-512 and M3D-128 organizations yield about 1.65% and 9.56% less average latency respectively. Shorter tRC time and shorter close-page access latencies for the M3D-512 and M3D-128 organizations result in lower CPI and average latency values for them.

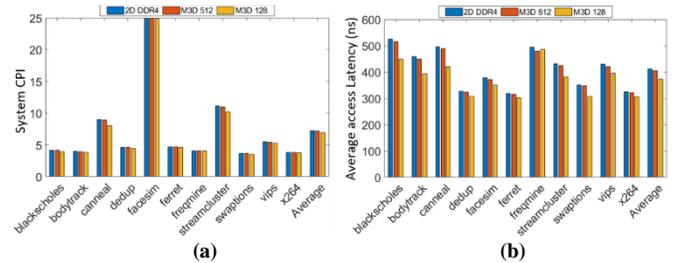

Fig. 9: (a) System cycles per instruction (CPI), and (b) average access latency results for the DDR4-512 (blue), M3D-512 (red), and M3D-128 (yellow) organizations across PARSEC benchmarks.

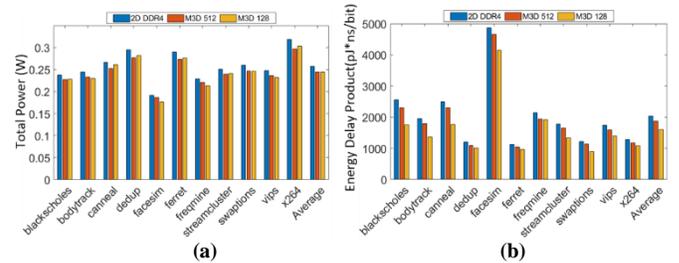

Fig. 10: (a) Total power, and (b) energy-delay product (EDP) results for the DDR4-512 (blue), M3D-512 (red), and M3D-128 (yellow) organizations across PARSEC benchmarks.

Fig. 10(a) shows total power values across all PARSEC benchmarks for our considered DRAM organizations. The total power is the sum of background power, activate power, burst power and refresh power. Compared to the baseline DDR4-512, M3D-512 and M3D-128 organizations yield on average about 4.97% and 4.96% lower power consumption respectively. The M3D organizations achieve these benefits with almost 14% less die size area. Fig. 10(b) shows energy-delay product (EDP) values. EDP indicates how balanced different designs are in terms of energy consumption and delay. We calculate EDP by multiplying energy per bit (pJ/bit) with average access latency (ns), while energy per bit is total power divided by throughput (bit/s). The results show that M3D-512 has

7.49% lower EDP than the baseline DDR4-512, whereas M3D-128 has 21.21% lower EDP than the baseline DDR4-512.

## VII. RELATED WORK

Despite being the most fundamental approach, having shorter bitlines is certainly not the only approach the researchers have explored for reducing the DRAM access latency. In fact, several other approaches have been reported in prior works that aim to optimize per-access as well as average DRAM latency. From many prior works available in the literature in this domain, we discuss only a few relevant prior works below. For example, prior works [1], [4], [5], [6], [16], [17], [18], and [19] they utilize different techniques to reduce the DRAM timing constraints and per-access latency. On the other hand, to improve average latency some other prior works (e.g., [28], [29]) aim to improve the access bandwidth of DRAMs by leveraging 3D-stacked DRAM structures.

Moreover, M3D technology is also an active research area. [3], [23], and [24] focus on design automation and layout generation for M3D integrated circuits and systems. Researchers have also used M3D integration technology for SRAM-based in-memory computing design in [20], [21], and [22]. Moreover, [26] has prototyped full M3D computing systems, and [27] presents the processor and memory integration using M3D technology. Further, [25] presents an MIVs based ultra-high-bandwidth on-chip memory bus design for M3D integrated DRAMs and non-volatile RAMs. In addition, in [40], M3D technology is explored as a possible solution to improve the performance of networks-on-chip.

## VIII. CONCLUSIONS

In this paper, we showed how the fundamental latency-area tradeoffs for DRAM can be mitigated by reorganizing DRAM cell-arrays using the emerging monolithic 3D (M3D) integration technology. We evaluated the latency-area tradeoffs for various configurations of 2D DDR4 and M3D DRAMs. Based on our evaluation results for PARSEC benchmarks, we found that our designed M3D DRAM cell-array organizations can yield up to 9.56% less latency, up to 4.96% less power consumption, and up to 21.21% less energy-delay product (EDP), with up to 14% less DRAM die area, compared to the conventional 2D DDR4 DRAM. These results corroborate the excellent capabilities of the M3D technology in mitigating the fundamental latency-area tradeoffs for DRAMs, to achieve simultaneous benefits in DRAM access latency and per-die cell density.